\documentclass[final,3p,times,twocolumn]{elsarticle}
 \biboptions{comma,sort&compress}
\usepackage{here}
 \usepackage{graphicx}
  \usepackage{epsfig}

 \usepackage{amssymb}
 \usepackage{amsthm}
\usepackage{amsmath}
\usepackage{amssymb}

\def\nin{\noindent}
\def\beq{\begin{equation}}
\def\eeq{\end{equation}}
\def\bea{\begin{eqnarray}}
\def\eea{\end{eqnarray}}

\newcommand{\T}[1]{\text{T} \left[ #1 \right]}

\newcommand{\GGVren}{\langle \hspace{-0pt} \frac{\alpha_s}{\pi} G^2 \hspace{-0pt} \rangle}
\newcommand{\GGMren}{\langle \hspace{-0pt} \frac{\alpha_s}{\pi} \hspace{-2pt} \left( \frac{(vG)^2}{v^2} - \frac{G^2}{4} \right) \hspace{-0pt} \rangle}

\journal{Nuc. Phys. (Proc. Suppl.)}

\begin{document}

\begin{frontmatter}



\title{In-Medium Modifications of Scalar Charm Mesons in Nuclear Matter}

 \author{T. Hilger, B. K\"ampfer}
 \address{Forschungszentrum Dresden-Rossendorf, PF 510119, D-01314 Dresden, Germany\\
TU Dresden, Institut f\"ur Theoretische Physik, 01062 Dresden, Germany}


\begin{abstract}
\nin
Employing QCD sum rules the in-medium modifications of scalar charm mesons
in a cold nuclear matter environment are estimated. The mass splitting
of $D^* - \bar D^*$ is quantified.
\end{abstract}


\end{frontmatter}


\section{Introduction}
\nin
In-medium modifications of hadrons, embedded in strongly interacting matter,
are of considerable interest, since they test the non-perturbative sector
of QCD. As density and temperature of the ambient medium can be changed
in a controllable manner, the changes of hadron properties, compared to
the vacuum, may serve as an in-depth check of our understanding of
hadron physics.

Previous investigations focussed on light vector mesons and light pseudoscalar
strange mesons \cite{Leupold_Metag_Mosel,Rapp_Wambach_Hees}, thus being
essentially restricted to the light-flavor ${\rm SU}_f(3)$ sector of QCD.
The starting FAIR project will enable the extension to charm degrees of freedom.
In fact, two major collaborations \cite{CBM,PANDA}
plan to investigate charm mesons and baryons in proton-nucleus,
anti--proton-nucleus and heavy-ion collisions.
Given this motivation, we are going to extend our recent study \cite{Hilger}
of the in-medium modifications of pseudoscalar open charm mesons to
the lowest excitations of scalar open charm mesons.
In doing so we employ QCD sum rules \cite{SVZ,SNB} and restrict ourselves to 
estimates of the mass splitting of scalar exciations which can be related to the currents
$\text{j}_{D^*} = \bar d c$ and $\text{j}_{\bar D^*} = \bar c d$
for small densities.

\section{QCD sum rules}
\nin
\nin
The Borel transformed in-medium sum rules for the current-current correlator
$\Pi(q) = i \int d^4x \, e^{iqx} \langle \langle \T{\text{j}(x) \text{j}^\dagger(0)} \rangle \rangle$
(here, $\T{\cdots}$ 
means the time-ordered product and 
$ \langle \langle \cdots \rangle \rangle$ stands for Gibbs average)
may be cast in the form
(cf.\ \cite{Hilger} for details)
\begin{eqnarray}
 && \left( \int_{\omega_0^-}^{\omega_0^+}
        + \int_{-\infty}^{\omega_0^-} + \int^{+\infty}_{\omega_0^+} \right)   d \omega \,
        \Delta \Pi(\omega,\vec{q}\,)
        \omega^{j} e^{-\omega^2/M^2} \nonumber \\
&& =  \pi {\cal B}_{Q^2\rightarrow M^2} \left[ \Pi^{j'}_{OPE}(Q^2,\vec{q}\,) \right] \left( M^2 \right) ,
\label{eq:even_b_sr}
\end{eqnarray}
where $Q^2 = -q_0^2$, $\Delta \Pi$ are the discontinuities along the entire real axis, $M$ is the Borel mass 
and the integral over the hadronic spectral function 
$\Pi(s,\vec{q}\,)$
is split into
a part of the low-energy excitations (first term) and the so-called continuum contributions (second and third terms).
The latter ones are mapped by a semi-local duality hypothesis 
to expressions corresponding to the operator product expansion (OPE)
of the correlator which yields ${\cal B} \left[ \Pi^{n'}_{OPE} \right]$.
Since particles and anti-particles behave differently in a nuclear medium at zero temperature,
two sum rules emerge -- an even one ($j=1$, $j' = e$) and an odd one
($j=0$, $j' = o$).    
The OPE's needed here can be obtained by combining the OPE's for pseudoscalar $D$ mesons in
\cite{Hilger} and the OPE's for difference sum rules in \cite{HilgerJPG}:
\begin{subequations} \label{eq:borel_sr_d}
\begin{align}
    & {\cal B}_{Q^2\rightarrow M^2} \left[ \Pi^e_{OPE}(Q^2, \vec{q} = 0\,) \right] \left( M^2 \right)
    \nonumber
    \\& \quad =
        \frac{1}{\pi} \int_{m_c^2}^\infty ds e^{-s/M^2} \text{Im} \Pi_{D^\ast}^{\rm per}(s, \vec{q} = 0\,)
        \nonumber
        \displaybreak[0]
        \\& \quad
        + e^{-m_c^2/M^2} \left(
        m_c \langle \overline{d}d \rangle
        - \frac{1}{2} \left( \frac{m_c^3}{2M^4} - \frac{m_c}{M^2} \right)
        \langle \overline{d}g\sigma{\cal G}d \rangle
        \right.
        \nonumber
        \displaybreak[0]
        \\& \quad
        + \frac{1}{12} \GGVren
        + \left[ \left( \frac{7}{18} + \frac{1}{3} \ln \frac{\mu^2 m_c^2}{M^4}
         - \frac{2\gamma_E}{3} \right)
        \left( \frac{m_c^2}{M^2} - 1 \right)
        \right.
        \nonumber
        \displaybreak[0]
        \\& \quad
        \left.
        - \frac{2}{3} \frac{m_c^2}{M^2}
        \right] \GGMren
        \nonumber
        \displaybreak[0]
        \\& \quad \left.
        + 2 \left( \frac{m_c^2}{M^2} - 1 \right) \langle d^\dagger iD_0 d \rangle
        \right.
        \nonumber
        \displaybreak[0]
        \\& \quad
        \left.
        - 4 \left( \frac{m_c^3}{2M^4} - \frac{m_c}{M^2} \right)
        \left[ \langle \overline{d} D_0^2 d \rangle
        - \frac{1}{8} \langle \overline{d} g \sigma {\cal G} d \rangle \right]
        \right) \: ,
\label{eq:borel_sr_d_even}
        \displaybreak[0]
    \\
    & {\cal B}_{Q^2\rightarrow M^2} \left[ \Pi^o_{OPE}(Q^2, \vec{q} = 0\,) \right] \left( M^2 \right)
            \nonumber
        \\& \quad =
        e^{-m_c^2/M^2} \left(
        \langle d^\dagger d \rangle
        - 4 \left( \frac{m_c^2}{2M^4} - \frac{1}{M^2} \right) \langle d^\dagger D_0^2 d \rangle
        \right.
        \nonumber
        \displaybreak[0]
        \\& \quad
        \left.
        - \frac{1}{M^2} \langle d^\dagger g \sigma {\cal G} d \rangle
        \right)
        \equiv e^{-m_c^2/M^2} \langle K(M) \rangle n,
\label{eq:borel_sr_d_odd}
\end{align}
\end{subequations}
where $\text{Im} \Pi_{D^\ast}^{\rm per}$ is given in \cite{SNB} and $m_c = 1.3$ GeV is the charm quark mass. 
We define 
$e \equiv \int_{\omega_0^-}^{\omega_0^+} d \omega \, \omega \, \Delta \Pi {\rm e}^{-\omega^2 /M^2}$ and
$o \equiv \int_{\omega_0^-}^{\omega_0^+} d\omega \, \Delta \Pi {\rm e}^{-\omega^2 /M^2}$
to obtain, with a pole ansatz for the lowest excitations 
within $\omega_- - \omega_+$, which couple with strenghts $F_\pm$  to the above currents,
\begin{subequations} \label{eq:mom_sys}
\begin{align}
e &= m_+ F_+ {\rm e}^{-m_+^2 / M^2}
   + m_- F_- {\rm e}^{-m_-^2 / M^2},\\
o &= F_+ {\rm e}^{-m_+^2 / M^2} -
     F_- {\rm e}^{-m_-^2 / M^2} .                       
\end{align}
\end{subequations}
The mass splitting $\Delta m$ and mass center $m$ are related by 
$m_\pm = m \pm \Delta m$. The following equations determine these quantities: 
\begin{subequations} \label{eq:m_sys}
\begin{align}
    \Delta m &= \frac12 \frac{ o e^\prime - e o^\prime}{ e^2 + o o^\prime } \: ,
    \label{eq:delta_m}
    \\
    m^2 &= \Delta m^2 - \frac{ e e^\prime + \left( o^\prime \, \right)^2 }
    { e^2 + o o^\prime }  \: ,
    \label{eq:m_cms}
\end{align}
\end{subequations}
where a prime denotes the derivative w.r.t. $M^{-2}$
and $e$ as well as $o$ are given by  (\ref{eq:borel_sr_d_even})
and (\ref{eq:borel_sr_d_odd}) minus the continuum parts from (\ref{eq:even_b_sr}).

\section{Low-density expansion}

An expansion in the density $n$ gives the leading term
for the mass splitting
\begin{align}
        \label{eq:dm1}
    \Delta m(n) & \approx \frac12 \frac{ \left. \frac{ { d} o }{ { dn} } \right|_{0}
        e^\prime(0) -  e(0) \bigl. \frac{ { d} o^\prime }{ { dn}} \bigr|_{0} }{ e(0)^2 }\, n
            \equiv \alpha_{\Delta m} n \: ,
\end{align}
since $e(n=0) \ne 0$ and $o(n=0) = 0$.

The primary goal of the present sum rule analysis is to find the dependence
of $\Delta m$ and $m$ on changes of the condensates entering (\ref{eq:borel_sr_d}).
However, also the continuum thresholds $\omega_0^\pm$ can depend on the density.
To study this influence, we consider the asymmetric splitting of the
continuum thresholds $\Delta \omega_0^2 = ((\omega_0^+)^2 - (\omega_0^-)^2)/2$ and parameterize its
density dependence by $\Delta \omega_0^2(n) = \alpha_{\Delta \omega} n + {\cal O}(n^2)$,
which leads to
\begin{subequations} \label{eq:ntlo_terms}
\begin{align}
    \left. \frac{ { d} o }{ { dn} } \right|_{0} =&
        \left( \frac{ e^{-\omega_0^2/M^2}}{\pi \omega_0} \text{Im} \Pi_{D^\ast}^{\rm per}(\omega_0^2) \frac{d \Delta \omega_0^2}{dn} \right)_{n=0}
        \nonumber
        \\
        & \quad
        + e^{-m_c^2/M^2} \langle K(M) \rangle
        \: ,
\end{align}
\begin{align}
    \left. \frac{ { d} o^\prime }{ { dn} } \right|_{0} =&
        \left( - \frac{ e^{-\omega_0^2/M^2}}{\pi} \omega_0\text{Im} \Pi_{D^\ast}^{\rm per}(\omega_0^2) \frac{d \Delta \omega_0^2}{dn} \right)_{n=0}
                \nonumber
        \\& \quad
        + e^{-m_c^2/M^2} \langle K^\prime(M) - m_c^2 K(M) \rangle \: .
\end{align}
\end{subequations}
The perturbative terms stem from the continuum contribution in case of unequal thresholds for particle and antiparticle.
In linear density approximation of the condensates the last terms become $o/n$ and $o^\prime/n$, respectively.
In this case, $\alpha_{\Delta m}$ is given as
\begin{equation}
\begin{split}
	\alpha_{\Delta m} = & -\frac{1}{2e(0)} \left(
		\frac on m^2(0) + \frac{o^\prime}{n}
		\right.
		\\
		& \left.
		+ \frac{e^{-\omega_0^2/M^2}}{\pi \omega_0} {\rm Im} \Pi_{D^\ast}^{\rm per} (\omega_0^2) \left[ m^2(0) - \omega_0^2\right] \alpha_{\Delta \omega}
		\right) \: ,
\end{split}
\end{equation}
which is dominated by the non-perturbative terms.
We choose the Borel mass range and the thresholds according to \cite{NDscalar}.

As an estimate for the order of $\alpha_{\Delta \omega}$ we rely on the splitting of the thresholds for the pseudoscalar channel in \cite{Hilger} and obtain
$\alpha_{\Delta \omega} \approx 0.25 \cdot 10^3$ GeV$^{-1}$.
It is an overestimation of the pseudoscalar ${\cal O}(n)$ threshold splitting as it would correspond to a linear interpolation of $\Delta \omega_0^2$ from the vacuum to nuclear saturation density and, hence, includes higher order terms in the density.
We choose $\alpha_{\Delta \omega} \approx 10^2 \ldots 10^3$ GeV$^{-1}$.

\begin{figure}[ht]
    \centering
            \includegraphics[width=0.4\textwidth]{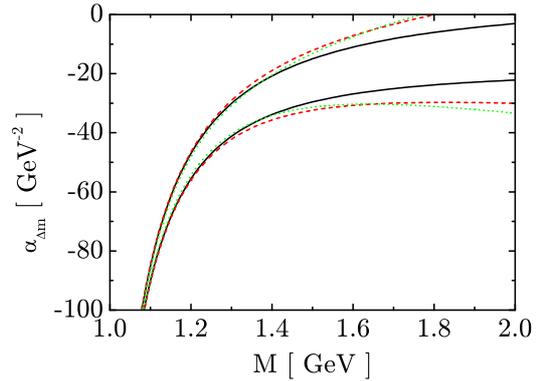}
    \caption{$\alpha_{\Delta m}$ as a function of the Borel mass for 
    $\alpha_{\Delta \omega}$ from  $-10^2 \text{ (lower bundle of curves)}$ GeV$^{-1}$ to $+10^2 \text{ (upper bundle of curves)}$ GeV$^{-1}$ 
and for threshold values $\omega_0^2 = 6.0$ GeV$^2$ 
    (solid black), 7.5 GeV$^2$ (dashed red) and 9.0 GeV$^2$ (dotted blue).}
    \label{fig1}
\end{figure}

\begin{figure}[ht]
    \centering
            \includegraphics[width=0.4\textwidth]{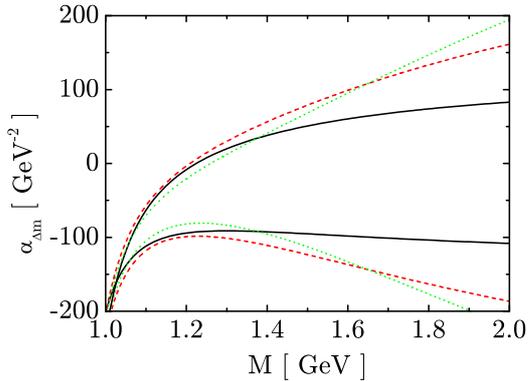}
    \caption{$\alpha_{\Delta m}$ as a function of the Borel mass for 
    $\alpha_{\Delta \omega}$ from  $-10^3 \text{ (lower bundle of curves)}$ GeV$^{-1}$ to $+10^3 \text{ (upper bundle of curves)}$ GeV$^{-1}$.
For line code see Fig.~\ref{fig1}.}
    \label{fig1b}
\end{figure}

\begin{figure}[ht]
    \centering
            \includegraphics[width=0.4\textwidth]{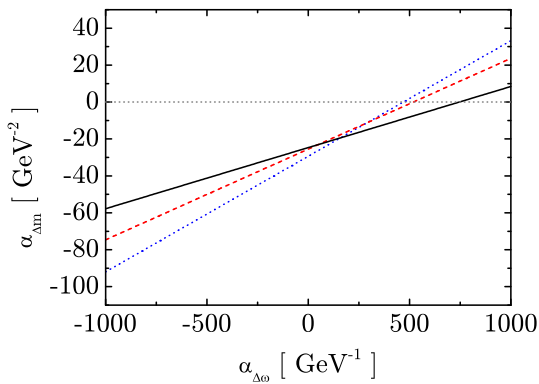}
    \caption{$\alpha_{\Delta m}$ as a function of $\alpha_{\Delta \omega}$ for $M=1.37$ GeV.
For line code see Fig.~\ref{fig1}.}
    \label{fig1c}
\end{figure}

The results are depicted in Fig.~\ref{fig1} for $\alpha_{\Delta \omega} = \pm 10^2$ GeV$^{-1}$
and for $\alpha_{\Delta \omega} = \pm 10^3$ GeV$^{-1}$ in Fig.~\ref{fig1b}.
In Fig.~\ref{fig1c} $\alpha_{\Delta m}$ as a function of $\alpha_{\Delta_\omega}$ for $M=1.37$ GeV, the minimum of the vacuum Borel curve for the scalar D meson, is displayed.

Considering the results for the mass splitting of heavy-light pseudoscalar mesons, e.g.\ $D$ and $B$, 
one could raise the question if the splitting is mainly caused by a splitting of the thresholds and, hence, 
might be an artifact of the method which determines $\Delta \omega_0^2$.
From the above study we find that $\alpha_{\Delta \omega}$ indeed influences the mass splitting.
A direct correlation in the sense of a correlation in sign can not be confirmed.
Furthermore, the results for $D_s$ mesons \cite{Hilger} allow for a positive mass splitting if the net strange quark density
 falls below a critical value.
As the strange quark density enters through the vector quark condensate, this already points to a suppressed 
influence of the threshold splitting on the mass splitting.

\section{Beyond low-density approximation}

In \cite{Hilger}, the threshold splitting is not considered as a free parameter but determined 
by the requirement that the minima of the Borel curves for particle and antiparticle 
are at the same Borel mass.

\begin{figure}[ht]
    \centering
            \includegraphics[width=0.4\textwidth]{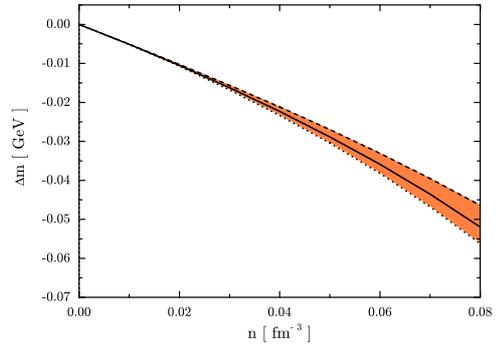}
    \caption{Mass splitting parameter $\Delta m$  of scalar $D^* - \bar D^*$ mesons 
    with the pole + continuum ansatz as a function of
density at zero temperature.
For a charm quark mass parameter of $m_c = 1.3$ GeV and mean threshold value $\omega_0^2 (0) = 7.5$ GeV$^2$.
    The curves are for
   $\omega_0^2(n) = \omega_0^2(0) + \xi n/n_0$ 
with $\xi = 0$ (solid), $\xi=1$ GeV$^2$  (dotted) and $\xi =-1$ GeV$^2$ (dashed).  }
    \label{fig2}
\end{figure}

\begin{figure}[ht]
    \centering
            \includegraphics[width=0.4\textwidth]{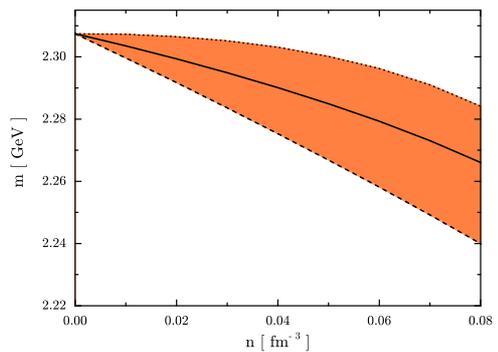}
    \caption{As in Fig.~\ref{fig2} but for the mean mass $m$.}
    \label{fig3}
\end{figure}

Following the same analysis strategy of \cite{Hilger} and employing  the condensates
listed in \cite{Hilger}, one obtains the
results exhibited in Figs.~\ref{fig2} and \ref{fig3}. 
This analysis goes beyond the strict linear density expansion of $m$ and $\Delta m$
and also takes into account quadratic terms in the density.
The medium dependent part of the chiral condensate 
(the density dependence of which is only in linear density approximation, as the other condensates too) 
enters the mass splitting next to leading order of the density.
The determination
of the mass center $m$ depends strongly on the chosen center of continuum thresholds $\omega_0^2 = ((\omega_0^+)^2 + (\omega_0^-)^2)/2$
as indicated by the broad range in Fig.~\ref{fig3} when varying their medium dependence.
In contrast, the splitting is fairly robust
as  evidenced by Fig.~\ref{fig1}, where the difference between curves of different thresholds is negligible.

A linear interpolation of the threshold splitting from vacuum to a density of $n=0.01 fm^{-3}$ gives an estimate for the ${\cal O}(n)$ term $\alpha_{\Delta \omega} \approx 7 \cdot 10^2$ GeV$^{-1}$, which justifies the range chosen in the previous section.

\section{Conclusions}
\nin
In summary we extend the recent analysis \cite{Hilger} of the QCD sum rules
to lowest scalar $D^*$ mesons. The importance of the density dependence
of the continuum thresholds is exposed. Going beyond the strict linear
approximation in density one obtains a robust pattern of the splitting of
scalar $D^* - \bar D^*$ mesons which resembles the one for pseudoscalars.
A firm prediction of the absolute values of the respective scalar
mesons is hampered by uncertainties in the determination of the mean
mass. With the employed values of the condensates entering
the truncated sum rule and within the employed analysis strategy
a tendency of a ''mass drop'' may be deduced.

\nin
For the sake of simplicity we restricted our analysis to a pole
+ continuum ansatz. One can go beyond such simplified treatment
of the hadronic spectral functions by considering moments.
The latter ones do not longer allow for a simple interpretation
but seem more appropriate for broad resonances. 
With respect to the research programme at FAIR, where precision
measurements of various charm hadrons are envisaged,
more model-independent studies are required.

\newpage

\section*{Acknowledgements}
\nin
The work is supported by GSI-FE and BMBF 06DR5059.

%













\end{document}